# An FE-EBC Method for Electromagnetic Scattering from Inhomogeneous Objects

Ehsan Khodapanah

*Abstract−* In this paper, we present a finite-element-extended boundary condition (FE-EBC) method for an efficient calculation of the electromagnetic wave scattering from inhomogeneous magneto-dielectric objects. To this end, we apply the hierarchical Legendre polynomial basis functions on large curved inhomogeneous hexahedral elements and propose an efficient numerical algorithm for a fast computation of the finite-element matrix entries in the presence of the material and Jacobian inhomogeneities. Also, we present a multiple-harmonic expansion on the surface boundary to represent the surface magnetic field accurately. The accuracy, efficiency and convergence of the method are studied through some numerical examples.

*Index terms−* Finite-element method (FEM), extended boundary condition (EBC) method, hexahedral elements, electromagnetic wave scattering, inhomogeneous objects.

## I. INTRODUCTION

The finite-element method (FEM) is a versatile and accurate numerical method for solving vector electromagnetic field problems, especially, the problem of electromagnetic wave scattering from arbitrary shape inhomogeneous scatterers [1-3].

In the case where the volume of the scatterer can be divided into a few large curved inhomogeneous hexahedral elements, the high order hierarchical version of the FEM becomes more preferable because of its excellent accuracy and convergence [4]. The main drawback of the method is that the calculation of the finite-element matrix entries in the presence of the material and Jacobian inhomogeneities is costly and time consuming. However, the universal array approach [5] and the product to some approach [6] can be utilized to accelerate the calculation of the matrix entries.

The main difficulty in the application of the FEM in solving infinite domain problems such as electromagnetic wave scattering problems is that the finite-element mesh should be truncated at a fictitious closed surface and an appropriate boundary condition, which is a mathematical equivalent to the sommerfeld radiation condition at infinity, should be imposed on the surface. The finite-element mesh can be truncated by applying an absorbing boundary condition (ABC) [7-11], which is an approximate and local boundary condition, or PMLs) [12-15], which extend the computational domain to the PML regions. An exact nonlocal boundary condition can be obtained by developing a surface integral equation on the fictitious surface through the application of the Green's functions [16-21]. When the fictitious surface for the truncation of the finite-element mesh belongs to a separable coordinate system (e.g., a spherical surface), the eigenfunctions of the exterior homogeneous domain can be used to develop an exact nonlocal boundary condition known as a Dirichlet to Neumann (DtN) boundary condition [22-27]. Application of the eigenfunctions of a separable system on a nonseparable fictitious surface or on another separable surface leads to the extended boundary condition (EBC) method [28-40]. Indeed, in the EBC method, a surface integral equation is imposed in the volume surrounded by the surface and is known as a null-field equation.

In this paper, we present a finite-element-EBC (FE-EBC) method to solve the electromagnetic wave scattering from inhomogeneous objects efficiently. To this end, a hierarchical hexahedral Legendre FEM is applied in the volume of the scatterer, along with a fast algorithm to calculate the volumetric finite-element matrix entries in the presence of the material and Jacobian inhomogeneities. Also, we present a multiple-harmonic expansion to represent the surface field very accurately.

---

The author is with the Faculty of Electrical and Computer Engineering, University of Tabriz, Tabriz, Iran (e-mail: ekhodapanah@tabrizu.ac.ir).

## II. EBC FORMULATION

The geometry of the problem is shown in Fig. 1, where the inhomogeneous magneto-dielectric object is surrounded by Γ and is illuminated by an incident electromagnetic wave. According to the equivalence principle [41], the surface current densities that correspond to the tangential scattered electromagnetic fields on Γ radiate a null field inside Γ. Mathematically, we have

$$j\omega\mu_0 \oint_\Gamma \bar{\mathbf{G}}(\mathbf{r},\mathbf{r}') \cdot (\hat{n} \times \mathbf{H}_s) ds' + \nabla \times \oint_\Gamma \bar{\mathbf{G}}(\mathbf{r},\mathbf{r}') \cdot (\hat{n} \times \mathbf{E}_s) ds' = 0 \ : \ \mathbf{r} \in \text{inside } \Gamma \quad (1)$$

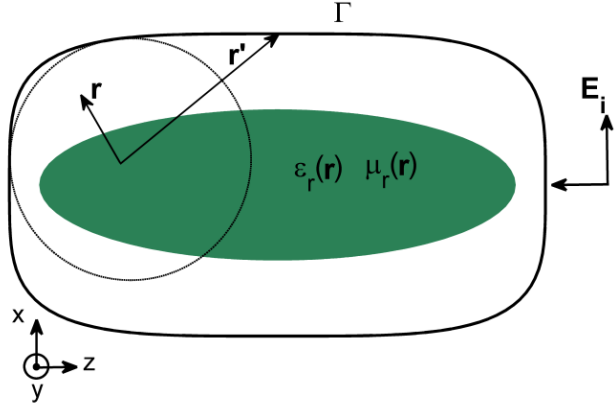

Fig. 1. Geometry of an inhomogeneous scatterer enclosed by Γ and illuminated by a uniform plane wave of the form $\mathbf{E}_i = \hat{x}\exp(jk_0 z)$.

where $\bar{\mathbf{G}}(\mathbf{r},\mathbf{r}')$ is a free-space dyadic Green's function [42] and $\mathbf{E}_s$ and $\mathbf{H}_s$ are the scattered electric and magnetic fields, respectively. The dyadic Green's function of the free space, $\bar{\mathbf{G}}(\mathbf{r},\mathbf{r}')$, can be expanded in terms of spherical vector wave functions as

$$\bar{\mathbf{G}}(\mathbf{r},\mathbf{r}') = -jk_0 \sum_i \boldsymbol{\psi}_{B,i}(k_0,\mathbf{r})\widehat{\boldsymbol{\psi}}_{H,i}(k_0,\mathbf{r}') = \frac{-j}{k_0}\sum_i \nabla \times \boldsymbol{\psi}_{B,i}(k_0,\mathbf{r})\nabla' \times \widehat{\boldsymbol{\psi}}_{H,i}(k_0,\mathbf{r}') \ : \ r < r' \quad (2)$$

where $\boldsymbol{\psi}$ is a divergence-free spherical vector wave function, $B$ and $H$ stand for Bessel and Hankel, respectively, and ^ is the complex conjugate of the angular part [42]. Substituting the first equality of (2) into the first integral in (1) and the second equality of (2) into the second integral in (1), we obtain a set of nonsingular integral equations as

$$-j\omega\mu_0 \oint_\Gamma \widehat{\boldsymbol{\psi}}_{H,i}(k_0,\mathbf{r}') \cdot (\hat{n} \times \mathbf{H}_s) ds' + \oint_\Gamma \nabla' \times \widehat{\boldsymbol{\psi}}_{H,i}(k_0,\mathbf{r}') \cdot (\hat{n} \times \mathbf{E}_s) ds' = 0 \quad (3)$$

where the coordinate origin to determine $\mathbf{r}'$ can be chosen as an arbitrary point inside Γ. The tangential scattered electric and magnetic fields on Γ can be expanded as

$$\hat{n} \times \mathbf{E}_s = \sum_c \sum_j A_j^c \, \hat{n} \times \boldsymbol{\psi}_{H,j}(k_0, \mathbf{r}_c') \quad (4)$$

$$\hat{n} \times \mathbf{H}_s = \sum_c \sum_j B_j^c \, \hat{n} \times \nabla_c' \times \boldsymbol{\psi}_{H,j}(k_0, \mathbf{r}_c') \quad (5)$$

Equations (4) and (5) represent the multiple-multipole field expansions on the surface with respect to several center points inside Γ, which are distinguished by different integers, $c$. Substituting (4) and (5) into (3) and using a manipulation similar to [42], we find that

$$A_j^c = -j\omega\mu_0 B_j^c \quad (6)$$

Equation (6) reveals the fact that the coefficients of the field expansions in (4) and (5) are related to each other. This fact will be used in the next section in the formulation of the FE-EBC method.

### III. FINITE-ELEMENT FORMULATION

Inside Γ, including inhomogeneities, the homogeneous vector wave equation holds as

$$\nabla \times \frac{1}{\mu_r(\mathbf{r})}\nabla \times \mathbf{E} - k_0^2 \varepsilon_r(\mathbf{r})\mathbf{E} = 0 \quad (7)$$

where $\varepsilon_r(\mathbf{r})$ and $\mu_r(\mathbf{r})$ are the relative permittivity and permeability of the scatterer bounded by Γ, respectively.

To discretize (7), we first expand the unknown vector electric field in terms of curl-conforming vector basis functions as

$$\mathbf{E} = \sum_j a_j \mathbf{U}_j \quad (8)$$

where $\mathbf{U}_j$'s are the vector basis functions and $a_j$'s are the unknown field coefficients. Substituting (8) into (7), taking the inner product of the both sides of (8) with the weighting functions $\mathbf{U}_i$'s (the Galerkin method), and applying the integration by parts on the first term, we arrive at

$$\sum_j a_j(S_{ij} - k_0^2 M_{ij}) - j\omega\mu_0 \oint_\Gamma \mathbf{U}_i \cdot \hat{n} \times (\mathbf{H}_s + \mathbf{H}_i)ds = 0 \quad (9)$$

where $\mathbf{H}_s$ and $\mathbf{H}_i$ are the scattered and incident magnetic fields, respectively, and $S_{ij}$ and $M_{ij}$ are the stiffness and mass matrix entries, respectively, which are defined as

$$S_{ij} = \int_V \frac{\nabla \times \mathbf{U}_i \cdot \nabla \times \mathbf{U}_j}{\mu_r} dv, \quad M_{ij} = \int_V \mathbf{U}_i \cdot \mathbf{U}_j \, \varepsilon_r \, dv \quad (10)$$

Substituting (5) into (9), we obtain an insufficient number of equations for the unknown coefficients $a_j$'s and $A_j$'s. To develop the complement equations, we make use of the continuity of the tangential electric field on Γ as

$$\hat{n} \times \mathbf{E} = \hat{n} \times \mathbf{E}_s + \hat{n} \times \mathbf{E}_i \quad \text{on } \Gamma \quad (11)$$

New equations can be obtained by taking the inner product of the both sides of (11) with a set of appropriate weighting functions. To make the final system symmetric, the weighting functions are chosen as the tangential scattered magnetic field bases in (5). By employing the method of section II, one can easily show that the (4) and (5) are still valid for the scattered electromagnetic fields on the surface of every sphere enclosing Γ. Therefore, once the system of equations are solved for $a_j$'s and $A_j$'s, the scattered fields can be calculated from (4) and (5).

In the following, we describe our efficient finite-element formulation for the filed inside Γ. The volume bounded by Γ is divided into a number of hexahedral elements. Each hexahedron is mapped onto a reference cubic element through an appropriate mapping. Then a set of hierarchical Legendre bases are defined for the field expansion in the reference element as

$$\begin{cases} P_m(u)P_{in}(v)P_{ip}(w)\,\hat{u} & \begin{cases} m = 0,1,\cdots,M-1 \\ n = 0,1,\cdots,N \\ p = 0,1,\cdots,P \end{cases} \\ P_{im}(u)P_n(v)P_{ip}(w)\,\hat{v} & \begin{cases} m = 0,1,\cdots,M \\ n = 0,1,\cdots,N-1 \\ p = 0,1,\cdots,P \end{cases} \\ P_{im}(u)P_{in}(v)P_p(w)\,\hat{w} & \begin{cases} m = 0,1,\cdots,M \\ n = 0,1,\cdots,N \\ p = 0,1,\cdots,P-1 \end{cases} \end{cases} \quad (12)$$

where $P_m$ is the $m$th degree Legendre polynomial and $P_{im}$ is the integrated Legendre polynomial, which is defined as

$$P_{im} = \begin{cases} \dfrac{P_0 - P_1}{2} & m = 0 \\ \dfrac{P_{m-1} - P_{m+1}}{4m + 2} & m = 1,2,\cdots M-1 \\ \dfrac{P_0 + P_1}{2} & m = M \end{cases} \quad (13)$$

The finite-element formulation of this paper is exactly the same as that has been used in [6] provided that we replace the Chebyshev polynomials of the first and second kinds in [6] with the integrated Legendre and Legendre polynomials, respectively. The numerical advantages that have been presented in [6] are also valid for the bases that have been defined in (12). For example, since the derivatives of the integrated Legendre polynomials are the Legendre polynomials, the derivative condition of [6] holds also for the basis functions in (12). This fact simplifies the computation of the stiffness matrix entries significantly [6]. Another advantage is the existence of an eight-fold symmetry in $[M_{uu}]$, $[M_{vv}]$, $[M_{ww}]$, $[F_1]$, $[F_2]$, and $[F_3]$ and a two-fold symmetry in $[M_{uv}]$, $[M_{uw}]$, $[M_{vw}]$, $[F_4]$, $[F_5]$, and $[F_6]$. Despite the aforementioned advantages, the required matrix entries should be calculated through direct 3D numerical integrations for general non-cubic non-homogeneous hexahedral elements, which are costly and time-consuming. In [6], a product to sum approach has been used for a fast and efficient computation of the matrix entries in a Chebyshev based FEM. In this paper, we present a very efficient numerical algorithm for a fast computation of the matrix entries in an arbitrary polynomial based FEM e.g., Chebyshev, Legendre, Jacobi, and even Lagrangian interpolating polynomial based FEMs. The proposed algorithm is very simple and its implementation in terms of computer programming is very easy. To describe the algorithm, we consider one of the mass submatrices e.g., $[M_{uu}]$ whose entries are defined as

$$M_{uu}^{m_1,m_2,n_1,n_2,p_1,p_2} = \iiint_{-1}^{1} P_{m_1}(u)P_{m_2}(u)Q_{n_1}(v)Q_{n_2}(v)\,R_{p_1}(w)R_{p_2}(w)\,\alpha(u,v,w)dudvdw \quad (14)$$

where $\alpha(u,v,w)$ is a product of the material and Jacobian inhomogeneities [6], and $P$, $Q$, and $R$ are arbitrary 1D polynomials in the $u$, $v$, and $w$ directions, respectively. The algorithm is as follows:

1) Calculate the 2D kernels $K^1_{p_1,p_2}(u,v)$ using the 1D integrals as

$$K^1_{p_1,p_2}(u,v) = \int_{-1}^{1} \alpha(u,v,w)\,R_{p_1}(w)R_{p_2}(w)dw \qquad p_1,p_2 = 0,1,\cdots,P \quad (15)$$

2) Calculate the 1D kernels $K^2_{n_1,n_2,p_1,p_2}(u)$ using the 1D integrals as

$$K^2_{n_1,n_2,p_1,p_2}(u) = \int_{-1}^{1} K^1_{p_1,p_2}(u,v)Q_{n_1}(v)Q_{n_2}(v)dv \qquad \begin{array}{l} p_1,p_2 = 0,1,\cdots,P \\ n_1,n_2 = 0,1,\cdots,N \end{array} \quad (16)$$

3) Calculate the submatrix entries using the 1D integrals as

$$M_{uu}^{m_1,m_2,n_1,n_2,p_1,p_2} = \int_{-1}^{1} K^2_{n_1,n_2,p_1,p_2}(u)P_{m_1}(u)P_{m_2}(u)du \qquad \begin{array}{l} p_1,p_2 = 0,1,\cdots,P \\ n_1,n_2 = 0,1,\cdots,N \\ m_1,m_2 = 0,1,\cdots,M-1 \end{array} \quad (17)$$

The number of multiply-add operations in the above algorithm is approximately

$$\frac{1}{2}(P+1)^2 s_1 s_2 2 s_3 + \frac{1}{4}(N+1)^2(P+1)^2 s_1 s_2 + \frac{1}{8}M^2(N+1)^2(P+1)^2 s_1 \quad (18)$$

where $s_1$, $s_2$, and $s_3$ are the number of quadrature points in the $u$, $v$, and $w$, respectively. The number of arithmetic operations per matrix entry is

$$s_1 + \frac{2 s_1 s_2}{M^2} + \frac{8 s_1 s_2 s_3}{M^2 (N+1)^2} \quad (19)$$

For an isotropic expansion ($M = N = P$ and $s_1 = s_2 = s_3$) and high degree polynomial bases (e.g., $M \geq 6$), an upper limit for (19) can be given as $s_1 + 19$. This means that every 3D integral of a matrix entry can be calculated at a cost of a 1D numerical integration.

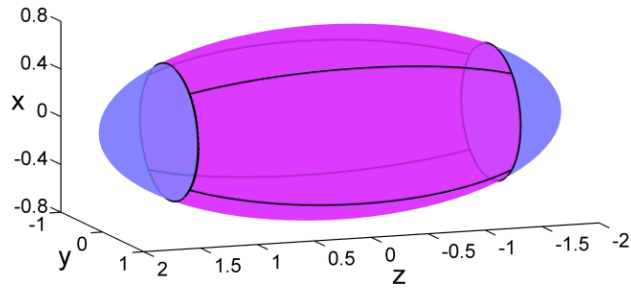

Fig. 2. Geometry of an ellipsoidal scatterer, which is modeled as a single curved hexahedral element. The parameters of the scatterer are $\varepsilon_r = 2$, $\mu_r = 1$, $a = 0.8$ m, $b = 1$ m, and $c = 2$ m.

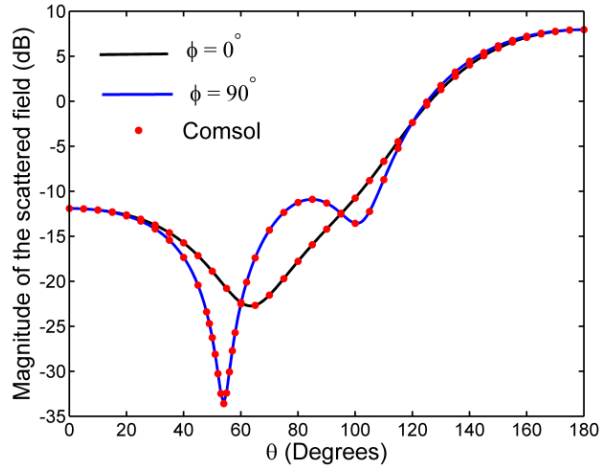

Fig. 3. Magnitude of the far zone scattered field ($r|\mathbf{E}_s|$) versus angle for the ellipsoid of Fig. 2 at $f = 100$ MHz.

Table I Average relative errors and computational times of the proposed method for the ellipsoid of Fig. 2.

| $[M,N,P,N_s]$ | Average relative error | Computational time (s) | | | | Number of harmonics on $\Gamma$ | DOF |
| --- | --- | --- | --- | --- | --- | --- | --- |
| | | Finite-element submatrix | Surface boundary submatrices | Matrix solving | total | | |
| $[4,5,7,2]$ | $2.8E-2$ | 0.4 | 0.8 | 0.1 | 1.3 | 48 | 650 |
| $[5,6,8,4]$ | $3.1E-3$ | 0.5 | 1.7 | 0.4 | 2.6 | 144 | 1119 |
| $[6,7,9,6]$ | $3.4E-4$ | 0.9 | 3 | 1.3 | 5.2 | 288 | 1762 |
| $[7,8,10,8]$ | $2.5E-5$ | 1.4 | 6 | 3.6 | 11 | 480 | 2597 |
| $[8,9,11,8]$ | $7.4E-6$ | 2.3 | 6.4 | 7.5 | 16.2 | 480 | 3402 |
| $[9,10,12,10]$ | $5.9E-7$ | 3.8 | 13.2 | 17.5 | 34.5 | 720 | 4627 |
| $[10,11,13,12]$ | $5.2E-8$ | 6 | 29 | 38 | 73 | 1008 | 6098 |
| $[11,12,14,14]$ | $2.6E-9$ | 9.5 | 56 | 77.5 | 143 | 1344 | 7833 |
| $[12,13,15,16]$ | — | 14 | 90 | 150 | 254 | 1728 | 9850 |
| Comsol | $4.7E-4$ | — | — | — | 285 | — | 445288 |

## IV. MULTIPLE-HARMONIC EBC METHOD

The FE-EBC formulation that has been derived in sections II and III can alternatively be obtained by applying the same finite-element formulation as before and substituting the surface magnetic field expansion, (5), and the surface component of the electric field that is obtained from the finite-element discretization of the volume into (3) to develop the complement equations. In this manner, the complement equations are obtained by imposing the null-field equation, (1), through (3) while (4) and (6) are unnecessary. In this viewpoint, the null-field condition is imposed in the volume of several inscribed spheres, which are centered at the center points of the multipole expansions. Unfortunately, the multiple-multipole expansion of the surface magnetic field in (5) converges slowly on $\Gamma$. To accelerate the convergence of the method, we apply a multiple-harmonic field expansion

$$\hat{n} \times \mathbf{H}_s = \sum_{c} \sum_{n=1}^{N_s} \sum_{m=-n}^{n} A_{nm}^c \, r_c' \, \nabla_c' Y_{nm}(\theta_c', \varphi_c') + B_{nm}^c \, \mathbf{r}_c' \times \nabla_c' Y_{nm}(\theta_c', \varphi_c') \quad (20)$$

where $Y_{nm}(\theta_c', \varphi_c')$ are scalar spherical harmonics as defined in [42].

Table II Convergence pattern of the FE-EBC method with different field expansions on $\Gamma$ for the Ellipsiod of Fig. 2 $M=12$, $N=13$, and $P=15$ are used for the inner FE solution

| Bessel based multiple-multipole expansion (as in the conventional EBC method) | | Multiple-harmonic expansion (Section IV) | | Hankel based multiple-multipole expansion (symmetric formulation of sections II and III) | |
| --- | --- | --- | --- | --- | --- |
| $N_s$ | Average relative error | $N_s$ | Average relative error | $N_s$ | Average relative error |
| 1 | $5.8E-1$ | 2 | $2.3E-2$ | 2 | $4.1E-2$ |
| 2 | $5.4E-3$ | 4 | $1.2E-3$ | 4 | $4.9E-3$ |
| 3 | $1.5E-3$ | 6 | $8.6E-5$ | 6 | $1.5E-3$ |
| 4 | $3.1E-5$ | 8 | $7.8E-6$ | 8 | $4.4E-4$ |
| 5 | $5.4E-6$ | 10 | $6.4E-7$ | 10 | $1.3E-4$ |
| 6 | $2.0E-6$ | 12 | $3.8E-8$ | 12 | $4.2E-5$ |
| 7 | $1.5E-5$ | 14 | $2.6E-9$ | 14 | $1.8E-5$ |
| 8 | $9.1E-6$ | 16 | $1.7E-10$ | 16 | $4.6E-6$ |
| 9 | $6.2E-5$ | 18 | — | 18 | $1.6E-6$ |

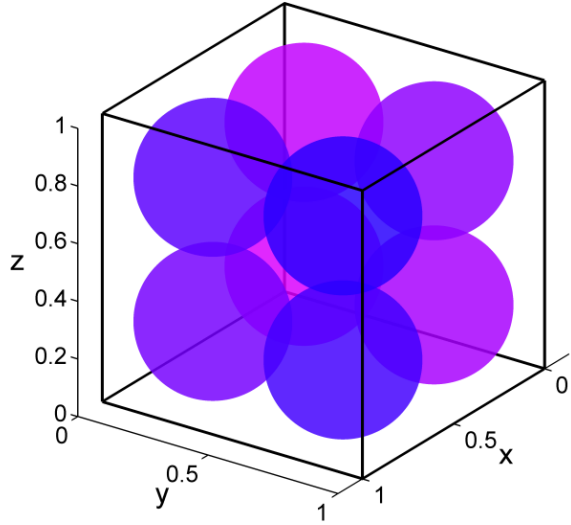

Fig. 4. Geometry of a unit inhomogeneous cubic scatterer. The parameters of the scatterer are $\varepsilon_r = 1 + \left(64xyz(1-x)(1-y)(1-z)\right)^3$ and $\mu_r = 1$.

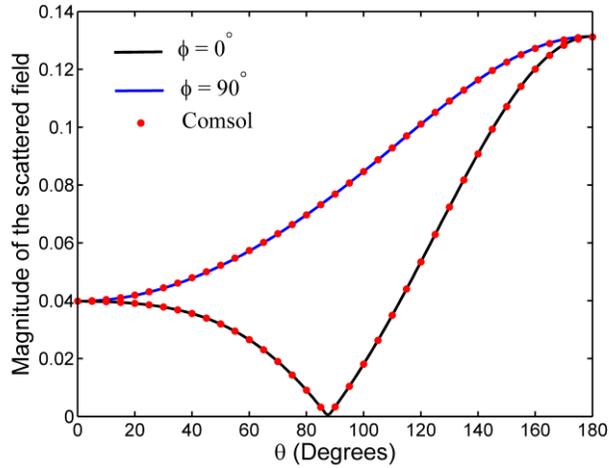

Fig. 5. Magnitude of the far zone scattered field ($r|\mathbf{E}_s|$) versus angle for the cubic scatterer of Fig. 4 at $f = 200$ MHz.

## V. NUMERICAL RESULTS

In this section, we investigate the efficiency, accuracy, and convergence of the proposed FE-EBC method through some numerical examples. All the calculations have been performed in a desktop computer equipped with an AMD Athlon II processor with a 3.2 GHz clock frequency and 8 GBs of RAM.

As the first example, we consider the electromagnetic wave scattering from an ellipsoidal object. The surface of the ellipsoid is defined as $(x/a)^2 + (y/b)^2 + (z/c)^2 = 1$. The parameters of the ellipsoid are $\varepsilon_r = 2$, $\mu_r = 1$, $a = 0.8$ m, $b = 1$ m, and $c = 2$ m. As shown in Fig. 2, the volume of the ellipsoid is modeled as a single curved hexahedral element. In order to model the curved surface of the ellipsoid exactly, we apply the following exact mapping for a cube to ellipsoid mapping

$$\begin{cases} x = a\,f(u,v)\sqrt{1-\dfrac{w^2}{2}} \\ y = b\,g(u,v)\sqrt{1-\dfrac{w^2}{2}} \qquad -1 \le u,v,w \le 1 \\ z = c\,w\sqrt{1-\dfrac{f^2+g^2}{2}} \end{cases} \qquad (21)$$

where

$$f(u,v) = u\cos\left(\frac{\pi}{4}v\right) + \sin\left(\frac{\pi}{4}u\right)$$
$$g(u,v) = v\cos\left(\frac{\pi}{4}u\right) + \sin\left(\frac{\pi}{4}v\right) - \frac{v}{\sqrt{2}}$$

The magnitude of the far zone scattered field $(r|\mathbf{E}_s|)$ at $f = 100$ MHz is shown in Fig. 3 as a function of angle and is compared with the Comsol multi-physics simulator. An excellent agreement is observed between the two results showing that the method is valid and accurate. In Fig. 2, $\Gamma$ is located at the surface of the ellipsoid and the center points for the multiple-harmonic expansion of the surface magnetic field on $\Gamma$ are located at the points $(0,0,-1.4)$, $(0,0,0)$, and $(0,0,1.4)$. The average relative errors over the angle and the computational times of the proposed method for the ellipsoid of Fig. 2 are listed in Table I for different values of $M$, $N$, $P$, and $N_s$. The table shows that the method converges rapidly and the computational time for the calculation of the finite-element submatrix, using the algorithm of section III, occupies a small portion of the total computational time, when the order of the bases increases.

In Table II, we compare the convergence pattern of our proposed multiple-harmonic expansion with the convergence patterns of conventional multiple-multipole expansions for the problem of Fig. 2. The table reveals the fact that the Bessel based multiple-multipole expansion does not converge properly due to the fact that, in this case, the condition number of the system matrix grows very rapidly and blocks the accuracy of the method. On the other hand, our proposed multiple-harmonic expansion converges properly on the number of harmonics and the condition number blockage is not seen in this case. Table II also shows that the Hankel based multiple-multipole expansion converges properly but its convergence rate is relatively slow in comparison with our proposed multiple-harmonic expansion.

As the second example, we consider the electromagnetic wave scattering from an inhomogeneous dielectric cube. The length of the cube is 1 m and $\mu_r = 1$ and $\varepsilon_r = 1 + \bigl(64xyz(1-x)(1-y)(1-z)\bigr)^3$. $\Gamma$ is located just at the surface of the cubic scatterer and the location of the inscribed spheres, inside which the null-field condition is imposed and around whose centers the multiple-harmonic expansion is applied, is shown in Fig. 4. The far zone scattered field magnitude versus angle at $f = 200$ MHz for the cubic scatterer of Fig. 4 is shown in Fig. 5 and is compared with Comsol. An excellent agreement is observed between the two results showing that the method is valid and accurate for the scatterer of Fig. 4. The convergence pattern and the computational times of the method in the calculation of the scattering from the cubic scatterer are shown in Table III for different values of $M$, $N$, $P$, and $N_s$. The table shows that when a non-smooth surface is used to truncate the finite-element mesh the number of harmonics on the surface will increase and the computational time for the calculation of the surface boundary submatrices will occupy a larger portion of the total computational time.

Finally, we consider the electromagnetic wave scattering from an inhomogeneous hemispheroidal magneto-dielectric object. The geometry of the hemispheroid and its model as a single curved hexahedral element is shown in Fig. 6. The geometry of the scatterer is specified by $(x/a)^2 + (y/a)^2 + (z/c)^2 = 1$, where $a = 1$ m and $c = 0.8$ m, and the electromagnetic properties of the scatterer are given by $\varepsilon_r = \exp((1.25z)^3)$ and $\mu_r = \sqrt{8z^3 + 1}$. The calculated scattered field as a function of angle at $f = 300$ MHz is depicted in Fig. 7 and is compared with Comsol. Excellent agreement is observed between the two results indicating that the proposed method is valid for inhomogeneous magneto-dielectric objects. By setting $M = 14$, $N = 14$, $P = 10$, and $N_s = 7$ the average relative error and the computational time of the proposed method are $2.9E - 5$ and 117 s, respectively, in a desktop computer whose characteristics have been mentioned at the beginning of this section.

Table III Average relative errors and computational times of the proposed method for the cubic scatterrer of Fig. 4.

| [$M, N, P, N_s$] | Average relative error | Computational time (s) | | | | Number of harmonics on $\Gamma$ | DOF |
|---|---|---|---|---|---|---|---|
| | | Finite-element submatrix | Surface boundary submatrices | Matrix solving | total | | |
| [5, 5, 5, 2] | $9.6E-3$ | 0.4 | 0.9 | 0.1 | 1.4 | 128 | 668 |
| [6, 6, 6, 2] | $1.7E-3$ | 0.55 | 1.3 | 0.3 | 2.15 | 128 | 1010 |
| [7, 7, 7, 3] | $4.4E-4$ | 0.85 | 2.5 | 1 | 4.35 | 240 | 1584 |
| [8, 8, 8, 3] | $5.8E-5$ | 1.4 | 2.6 | 2.3 | 6.3 | 240 | 2184 |
| [9, 9, 9, 5] | $5.8E-6$ | 2.2 | 8.1 | 6.7 | 17 | 560 | 3260 |
| [10, 10, 10, 7] | $3.8E-7$ | 3.5 | 22.5 | 17.5 | 43.5 | 1008 | 4638 |
| [11, 11, 11, 8] | $7.3E-8$ | 5.7 | 38.5 | 37 | 81.2 | 1280 | 6032 |
| [12, 12, 12, 10] | $-$ | 8.5 | 92.5 | 83 | 184 | 1920 | 8004 |
| Comsol | $1.7E-3$ | $-$ | $-$ | $-$ | 182 | $-$ | 330525 |

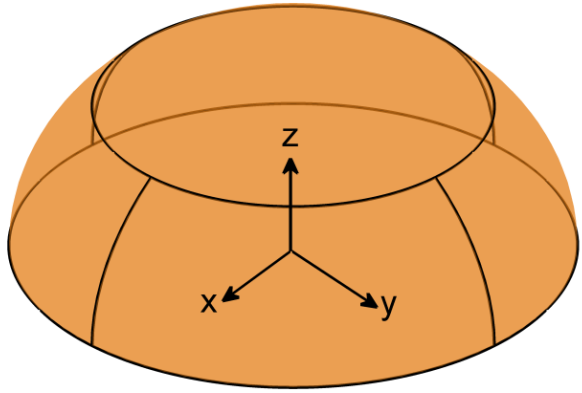

Fig. 6. Geometry of an inhomogeneous hemispheroidal magneto-dielectric scatterer, which is modeled as a single curved hexahedral element. The parameters of the scatterer are $\varepsilon_r = \exp((1.25z)^3)$, $\mu_r = \sqrt{8z^3 + 1}$, $a = 1$ m, and $c = 0.8$ m.

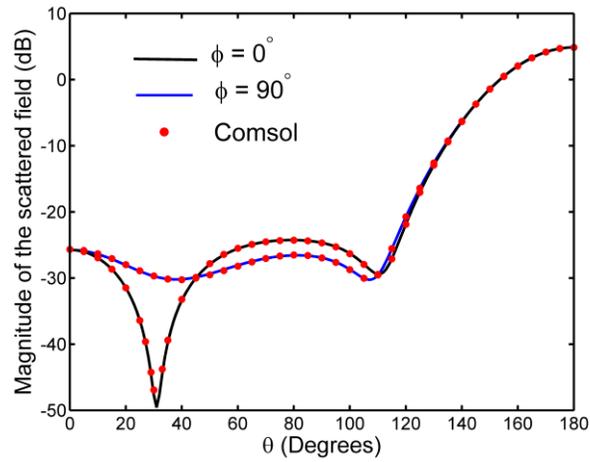

Fig. 7. Magnitude of the far zone scattered field ($r|\mathbf{E}_s|$) versus angle for the hemispheroid of Fig. 6 at $f = 300$ MHz.

## VI. Conclusion

An FE-EBC method has been proposed for the electromagnetic wave scattering from inhomogeneous objects. An algorithm has been proposed for a fast computation of the volumetric finite-element matrix entries and a multiple-harmonic expansion has been applied for an accurate representation of the magnetic field on the surface. The method has been applied to calculate the electromagnetic wave scattering from some dielectric scatterers. The results have shown that the method converges rapidly and the computational time is relatively small compared with the other FEM based solvers.

## References

1. J. M. Jin, *The Finite Element Method in Electromagnetics,* New York: John Wiley & Sons, 2002.
2. J. L. Volakis, A. Chatterjee, and L. C. Kempel, *Finite Element Method for Electromagnetics*, Piscataway, NJ: IEEE Press, 1998.
3. P. P. Silvester and R. L. Ferrari, *Finite Elements for Electrical Engineers*, 3rd ed. Cambridge, U.K.: Cambridge Univ. Press, 1996.
4. M. M. Ilic and B. M. Notaros, "higher order hierarchical curved hexahedral vector finite elements for electromagnetic modeling," *IEEE Trans. Microw. Theory Techn.,* vol. 51, no. 3, pp. 1026-1033, Mar. 2003.
5. D. Ansari Oghol Beig, J. Wang, Z. Peng, and J. F. Lee, "A universal array approach for finite elements with continuously inhomogeneous material properties and curved boundaries," *IEEE Trans. Antennas Propag.,* vol. 60, no. 10, Oct. 2012.
6. E. Khodapanah, "Efficient computation of the matrix entries in a hierarchical-hexahedral finite-element solution of curl-curl equation," *IEEE Trans. Microw. Theory Techn.,* vol. 66, no. 6, pp. 2697-2703, April 2018.
7. B. Engquist, and A. Majda, "Absorbing boundary conditions for the numerical simulation of waves," *Math. Comput.*, vol. 31, pp. 629-651, 1977.
8. A. Bayliss, and E. Turkel, "Radiation boundary conditions for wave-like equations," *Commun. Pure Appl. Math.*, vol. 33, pp. 707-725, 1980.
9. A. F. Peterson, "Absorbing boundary conditions for the vector wave equation," *Microwave Opt. Tech. Lett.*, vol. 1, pp. 62-64, 1988.
10. J. P. Webb and V. N. Kanellopoulos, "Absorbing boundary conditions for the finite element solution of the vector wave equation," *Microwave Opt. Tech. Lett.* , vol. 2, 370-372, 1989.
11. J. D'Angelo, and I. D. Mayergoyz, "Finite element methods for the solution of RF radiation and scattering problems," *Electromagnetics*, vol. 10, pp. 177-199, 1990.
12. J. P. Berenger, "A perfectly matched layer for the absorption of electromagnetic waves," *J. Comput. Phys.*, vol. 114, pp. 185-200, 1994.
13. C. Chew, and W. H. Weedon, "A 3D perfectly matched medium from modified Maxwell's equations with stretched coordinates," *Microwave Opt. Tech. Lett.*, vol. 7, pp. 599-604, 1994.
14. Z. S. Sacks, D. M. Kingsland, R. Lee, and J. F. Lee, "A perfectly matched anisotropic absorber for use as an absorbing boundary condition," *IEEE Trans. Antennas Propagat.*, vol. 43, pp. 1460-1463, 1995.
15. J. P. Berenger, "Three-dimensional perfectly matched layer for the absorption of electromagnetic waves," *J. Comput. Phys.*, vol. 127, pp. 363-379, 1996.
16. B. H. McDonald, and A. Wexler, "Finite-element solution of unbounded field problems," *IEEE Trans. Microwave Theory Tech.*, vol. 20, pp. 841-847, 1972.
17. K. D. Paulsen, D. R. Lynch, and J. W. Strohbehn, "Three-dimensional finite, boundary, and hybrid element solutions of the Maxwell equations for lossy dielectric media," *IEEE Trans. Microwave Theory Tech.*, vol. 36, pp. 682-693, 1988.
18. X. Yuan, "Three-dimensional electromagnetic scattering from inhomogeneous objects by the hybrid moment and finite element method," *IEEE Trans. Microwave Theory Tech.*, vol. 38, pp. 1053-1058, 1990.
19. J. M. Jin, and J. L. Volakis, "Electromagnetic scattering by and transmission through a three-dimensional slot in a thick conducting plane," *IEEE Trans. Antennas Propagat.*, vol. 39, pp. 543-550, 1991.
20. J. M. Jin, J. L. Volakis, and J. D. Collins, "A finite element-boundary integral method for scattering and radiation by two and three-dimensional structures," *IEEE Antennas Propagat. Mag.*, vol. 33, no. 3, pp. 22-32, 1991.
21. J. J. Angelini, C. Soize, and P. Soudais, "Hybrid numerical methods for harmonic 3-D Maxwell equations: Scattering by a mixed conducting and inhomogeneous anisotropic dielectric medium," *IEEE Trans. Antennas Propagat.*, vol. 41, pp. 66-76, 1993.
22. S. K. Jeng, and C. H. Chen, "On variational electromagnetics: Theory and application," *IEEE Trans. Antennas Propagat.*, vol. 32, pp. 902-907, 1984.
23. R. B. Wu, and C. H. Chen, "Variational reaction formulation of scattering problem for anisotropic dielectric cylinders," *IEEE Trans. Antennas Propagat.*, vol. 34, 640-645, 1986.

24. J. F. Lee, and Z. J. Cendes, "Transfinite elements: A highly efficient procedure for modeling open field problems," *J. Appl. Phys.*, vol. 61, pp. 3913-3915, 1987.
25. J. B. Keller, and D. Givoli, "Exact non-reflecting boundary conditions," *J. Comput. Phys.*, vol. 82, pp. 172-192, May 1989.
26. P. C. Allilomes, and G. A. Kyriacou, "A nonlinear finite-element leaky waveguide solver," *IEEE Trans. Microw. Theory Tech.,* vol. 55, no. 7, pp. 1496-1510, Jul. 2007.
27. E. Khodapanah, "Numerical separation of vector wave equation in a 2-D doubly-connected domain," *IEEE Trans. Microw. Theory Techn.,* vol. 62, no. 11, pp. 2551-2562, Nov. 2014.
28. P. C. Waterman, "Matrix Formulation of Electromagnetic Scattering," *Proc. IEEE*, vol. 53, pp. 805-811, 1965.
29. R. H. T. Bates," Modal expansions of electromagnetic scattering from perfectly conducting cylinders of arbitrary cross-section," *Proc. IEE*, vol. 115, no. 10, pp. 1443-1445, 1968.
30. P. C. Waterman, "New Formulation of Acoustic Scattering," *J. Acoust. Soc. Am.*, vol. 45, pp. 1417-29, 1969.
31. P. C. Waterman, "Symmetry, Unitarity, and Geometry in Electromagnetic Scattering," *Phys. Rev. D*, vol. 3, pp. 825-829, 1971.
32. J. C. Bolomey, and A. Wirgin, "Numerical comparison of the Green's function and the Waterman and Rayleigh theories of scattering from a cylinder with arbitrary cross-section," *Proc. IEE*, vol.121, no. 8, pp. 794-804, 1974.
33. P. W. Barber, and C. Yeh, "Scattering of electromagnetic waves by arbitrary shaped dielectric bodies," *Appl. Opt.*, vol. 14, pp. 2864-2872, 1975.
34. R. H. T. Bates, and D. J. N. Wall, "Null field approach to scalar diffraction, I. General methods, II. Approximate methods," *Philos. Trans. R. Soc. Londo Ser. A*, vol. 287, pp. 45-95, 1977.
35. A. G. Ramm, "Convergence of the T-matrix approach to scattering theory," *J. Math. Phys.*, vol. 23, pp. 1123-25, 1982.
36. M. F. Iskander, A. Lakhtakia, and C. H. Durney, "A new iterative procedure to solve for scattering and absorption by lossy dielectric objects," *Proc. IEEE*, vol. 70, pp. 1361-1362, 1982.
37. M. F. Iskander, A. Lakhtakia, and C. H. Durney, "A new procedure for improving the solution stability and extending the frequency range of the EBCM," *IEEE Trans. Antennas Propagat.*, vol. AP-31, pp. 317-324, 1983.
38. A. Bostrom, "Scattering of acoustic waves by a layered elastic obstacle in a fluid-An improved null field approach," *J. Acoust. Sot. Am.*, vol. 76, pp. 588-593, 1984.
39. R. H. Hackman, "The transition matrix for acoustic and elastic wave scattering in prolate spheroidal coordinates," *J. Acoust. Soc. Am.*, vol. 75, pp. 35-45. 1984.
40. A. Doicu, and T. Wriedt, "Extended boundary condition method with multipole sources located in the complex plane," *Optics. Comm.*, vol. 139, pp. 85-91, 1997.
41. R. F. Harrington, *Time-harmonic electromagnetic fields*. New York: McGraw-Hill, 1961.
42. W. C. Chew, *Waves and Fields in Inhomogeneous Media,* New York: IEEE Press, 1995.